# The evolution of cooperation in the public goods game on the scale-free community networks under multiple strategy updating rules

Mingzhen Zhang, Naiding Yang, Xianglin Zhu

School of Management, Northwestern Polytechnical University, Xi'an 710000, China

Abstract:

Social networks have a scale-free property and community structure, and many problems in life have the characteristic of public goods, such as resource shortage. Due to different preferences of individuals, there exist individuals who adopt heterogeneous strategies updating rules in the network. We investigate the evolution of cooperation in the scale-free community network with public goods games and the influence of multiple strategy updating rules. Here, two types of strategy updating rules are considered which are pairwise comparison rules and aspiration-driven rules. Numerical simulations are conducted and presented corresponding results. We find that community structure promotes the emergence of cooperation in public goods games. In the meantime, there is a "U" shape relationship between the frequency of cooperators and the proportion of the two strategy updating rules. With the variance in the proportion of the two strategy updating rules, pairwise comparison rules seem to be more sensitive. Compared with aspiration-driven rules, pairwise comparison rules play a more important role in promoting cooperation. Our work may be helpful to understand the evolution of cooperation in social networks.

Keywords: Community structure; Public goods game; Cooperation; Multiple strategy-updating rules

## 1. Introduction

With the rapid economic development, human beings are gradually facing more and more environmental problems while enjoying the convenience brought by technology, such as resource shortage, climate warming, serious pollution, etc. These problems are typical social dilemmas where individual interests conflict with collective interests, and individuals choose to betray the collective for the sake of getting higher self-interest. However, cooperative behavior seems to widely exist in living systems, from microbial groups[1][2], social animals[3][4] to human society[5]. Hence, the emergence and maintenance of cooperation remain to be an open question. Aiming to resolve such a problem, the evolutionary game theory has been proposed as a meaningful and powerful framework. Theoretical analyses are mainly performed on the three typical two-player games:1) prisoner's dilemma game (PDG); 2) snowdrift game (SDG); and 3) the stag-hunt game (SHG)[6][7][8]. These three simple games are usually used to explain cooperative or uncooperative behavior through pairwise interactions. To provide a reasonable explanation for group interactions beyond pairwise interactions, the public goods game (PGG) has been proposed as a more general model. PGG as an expansion of PDG is a typical model to describe group interaction, which can effectively address the problem of global warming, resource shortage, and team cooperation. In the PGG model, cooperative individuals input resources into the public pool, but the defective individuals pay nothing. And the final payoff is divided equally among all individuals. Thus, defecting is the dominating strategy for individuals in the PGG model, but cooperative behavior exists widely in the social and biology world. Therefore, various effective mechanisms have been proposed to understand the emergence of cooperation.

Nowak[9] firstly summarized five mechanisms for the evolution of cooperation, which are kin selection, direct reciprocity, indirect reciprocity, group selection, and network reciprocity. Among them, the network reciprocity mechanism has attracted a great deal of attention. Because individuals interact more in the spatial structure or social network, when the cluster of cooperators outcompetes the defectors, it will promote the emergence of cooperation. The seminal work of Nowak and May[10], which found that spatial structure induces the emergence of cooperation, has opened the research of evolutionary game on the network. Inspired by this idea, research on PGG on complex networks is springing up. Many researchers have studied the evolutionary dynamics of the PGG on lattice structure. Since real-world systems are often heterogenous, the regular lattice structure is far from describing its topology. Nowadays, the research objective is shifting from evolutionary games on the regular lattice to evolutionary games on social networks[11][12][13]. One of the realistic structures in social networks is the scale-free property, another is the community structure[14]. In the community network, nodes within the same community are closely connected, while nodes between different communities are sparsely connected[15]. The community structure represents different meanings in the real world. In social networks, different communities represent different fields, occupations, ages, etc. in citation networks, different communities represent different research fields. Based on this, scholars have introduced the network have both scale-free property and community structure[16]. There has been a variety of research on PGG on the scale-free network[17][18][19], but there is a lack of research about PGG on community networks.

Apart from the research on the relationship between network structure and the evolution of cooperation, introducing strategic complexity is another way of bringing the public goods game closer to reality. Many researchers have paid attention to the effect of heterogeneous strategy updating rules on the evolution of cooperation. From the perspective of population structure, the existing research can be divided into well-mixed networks and heterogeneous networks. Liu (2015)[20] combined the Moran process and Pairwise comparison rules into the stochastic evolutionary game with a well-mixed finite population. In the well-mixed population, there are M individuals who take the Moran process and N individuals who take the Pairwise comparison rules in the updating process. After a series of deductions, the expressions of fixation times and fixation probabilities[21] can be obtained. Then the heterogeneous strategy updating rules on the spatial evolutionary game have been investigated. Quan and You have studied the relationship between multi-strategy updating rules and the evolution of cooperators on the heterogeneous network. Quan (2020)[22] considers the different updating rules as different pieces of evidence and introduces evidential reasoning theory to construct a new strategy updating rules. He integrates the two aspects of information from both imitation-based updating rules and aspiration-based updating rules as a new updating rule into spatial public goods game on a square lattice. The simulation results show that evidential reasoning can effectively promote the emergence of cooperation under appropriate values of fusion weight. You (2020)[23] investigate the effect of the combination of strategy learning mechanism and strategy teaching mechanism on cooperation in prisoner's dilemma on a regular lattice. He thoroughly studied the relationship between the proportional coefficient of the two updating rules and the frequency of cooperators. After that, You T (2021)[24] further analyses the mutation mechanism in the heterogeneous strategy updating rules. At the end of each generation, individuals can transform their strategy updating rules under certain conditions. The effect of population proportion and individual rule mutation rate on the emergence of cooperation have been investigated. Zou (2020)[25] studied the PDG model in a lattice network in which partial individuals update their strategies by imitating their neighbors' strategies and remaining individuals update their strategies based on aspiration level. The above research all investigates the effect of strategy updating rules on the evolution of cooperation on square lattice networks. In order to describe the evolution of cooperative behaviors in realistic network, it is necessary to investigate the effect of multiple strategy updating rules on the emergence of cooperation in scale-free network community network.

Given the scale-free property and community property of social networks, this paper generates a scale-free network with the characteristics of community structure and explores how cooperative behaviors evolve where there exist multiple strategy updating rules in the public goods game on the network. The rest of the paper is organized as follows. Section 2 demonstrates the generation of the scale-free community network, the public goods game model and the strategy-updating rules. Section 3 discusses the results from numerical simulations in detail. The main work and conclusions are presented in section 4 at last.

## 2. Model

### 2.1 Public goods game on the scale-free community network

#### 2.1.1 Constructing the scale-free network with community structure

The scale-free network with community structure is generated according to the algorithm proposed by Li (2005)[26] and Liu (2017)[27]. In a community network, the edges among the nodes in the same community are denser, whereas the edges among the nodes in different communities are sparser. The strength of the community structure can be quantified by using a modularity measure, which for a random network will be close to zero and will be close to one for a strong community structure.

We consider a complex product network consisting of M well-mixed communities, each community has an equal size N. Let $q$ denote the probability of nodes crossing communities, which represents the degree of interaction between communities.

#### 2.1.2 Public goods game

A social dilemma is often expressed as a multi-person dilemma, thus public goods game as the typical game model used to describe the group interaction has gotten a great deal of attraction recently. In the public goods game, individuals have two strategies to choose, which are cooperate (C) and defect (D). Cooperators contribute all resources to the public goods, but defectors put nothing into the public goods. The total resource is multiplied by an enhancement factor r and is redistributed equally among all individuals in the group. And the resource of every individual is set to be fixed, which is 1.

In the public goods game, each individual $i$ will participate in $k_i + 1$ public goods games centered on individual $i$ and its neighbors. Each public goods game is composed of a central individual and its neighbors.

$$\pi_i = \sum_{l \in \Omega_i} s_l \cdot \frac{1}{k_l}$$

$$\Pi_i = \pi_i - 1$$

### 2.2 Multiple strategy-updating rules

The strategy updating rules used in network evolutionary games generally include pairwise comparison[28] and aspiration-driven rules[29][30][31]. Here we suppose that there coexist the two updating rules in the network and investigate how the cooperative behavior evolve when the proportion of the two updating rules change. Then we introduce updating rules as follows:

#### 2.2.1 Pairwise comparison rules

An individual randomly chooses one neighbor and compares his profits with the neighbor. And the probability that he will imitate his neighbor's strategy in the next round of the game is proportional to the difference between their profits. This updating rule can be expressed as follows:

$$W(s_i \leftarrow s_j) = \frac{1}{1 + exp[(\Pi_i - \Pi_j)/\sigma]}$$

where $W(s_i \leftarrow s_j)$ represents the probability of individual i imitates the strategy of individual j. $s_i$ and $s_j$ represent the strategy of individual i and individual j, respectively. $\Pi_i$ and $\Pi_j$ represent the profit of individual *i* and individual j, respectively. $\sigma$ represents the bounded rationality or noise in the process of strategy adoption[32].

2.2.2 Aspiration-driven rules

An individual will compare its profit with its aspiration level, the individual *i* will change its strategy in the next round of game with the probability $W(s_i \leftarrow \bar{s}_i)$, which is proportional to the difference between its profit and its aspiration level.

Chen[33] introduced a parameter A that indicates the aspiration level of players, and each player calculates its aspiration payoff based on parameter A. The aspiration payoff of player i is $\rho_i = Ak_i$, with $k_i$ denotes the number of neighbors of individual i. In [34], the aspiration level is proportional to the degree of nodes. Zhang (2020)[31] supposed that the aspiration level of all individuals is equal. The above researches consider the position of nodes, but they ignore that how much payoff can these nodes make.

In this paper, we deem that the aspiration level should be proportional to the degree of nodes and the payoff which they make. Thus, the aspiration level $\rho_i$ is expressed as follows:

$$\rho_i = \frac{k_i}{\sum_{j \in \Omega_i} k_j} \cdot \sum_{j \in \Omega_i} \Pi_j$$

The aspiration-driven updating rules is expressed as follows:

$$W(s_i \leftarrow \bar{s}_i) = \frac{1}{1 + \exp[(\Pi_i - \rho_i)/\sigma]}$$

where $W(s_i \leftarrow \bar{s}_i)$ represents the probability of individual I changes its strategy. $\rho$ represents the aspiration level of individual i. $\Pi_i$ represent the profit of individual I, and $\sigma$ represents the bounded rationality or noise in the process of strategy adoption.

Here, we suppose that there coexist three strategy updating rules.

3. Numerical simulations

To illustrate the effects of multi-strategy updating rules on the evolution of cooperation in the public goods game on community network, several numerical simulations and corresponding discussions are presented.

3.1 The scale-free network with community structure

To explore the impact of community structure on cooperation, we firstly introduced some concepts about community networks. One of the concepts is proposed by Newman[34] in 2004 to evaluate community structure in the network, which is called modularity and can be expressed as follows:

$$Q = \frac{1}{2m} \sum_{ij} \left[ A_{ij} - \frac{k_i k_j}{2m} \right] \delta(c_i, c_j)$$

where $m$ represents the number of all the links in the network, $A_{ij}$ represents the adjacency matrix of the network, $k_i$ represents the degree of node i and $c_i$ represents the community where node *i* belongs to and $\delta(c_i, c_j)$ is a segmentation function. If $c_i = c_j$, then $\delta(c_i, c_j) = 1=$; otherwise, $\delta(c_i, c_j) = 0$.

The larger the Q, the better the effect of community division. The value of Q ranges from -0.5 to 1

generally. Newman points out that when the value of Q is between 0.3 and 0.7, the effect of clustering is well.

Another concept is connection probability. The connection probability is proposed in the community network generation algorithm, which means the probability that the new node connects with nodes in other communities.

Following the study of Li (2005)[26] and Liu(2017)[27], we construct the network with 500 nodes and four communities. The connection probability $q$ is an adjustable variable. By adjusting the value of connection probability, we can obtain many community networks with different modularity. The results are shown in Table 1 and Fig.1.

Table 1. The relationship between connection probability and modularity

| $q$ | 0     | 0.05  | 0.1   | 0.15  | 0.2   | 0.25  | 0.3   | 0.35  | 0.4   | 0.45  |
|-----|-------|-------|-------|-------|-------|-------|-------|-------|-------|-------|
| Q   | 0.824 | 0.801 | 0.799 | 0.776 | 0.767 | 0.732 | 0.704 | 0.709 | 0.685 | 0.673 |
| $q$ | 0.5   | 0.55  | 0.6   | 0.65  | 0.7   | 0.75  | 0.8   | 0.85  | 0.9   | 0.95  |
| Q   | 0.644 | 0.632 | 0.627 | 0.604 | 0.602 | 0.585 | 0.563 | 0.56  | 0.547 | 0.542 |

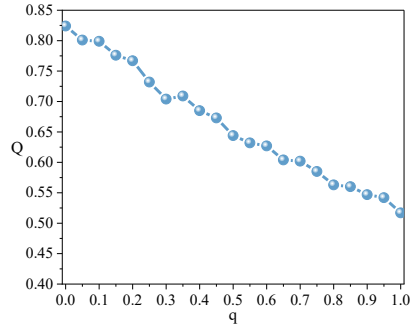

**Fig.1.** Relationship between connection probability q and modularity Q. All the results are based on N=500, m/n=2.

As shown in Fig 1., we can find that when the value of connection probability is bigger, the value of modularity gets smaller. For convenience, we take connection probability $q$ to measure the characteristics of community structure in the following simulations. The simulation environment is under network size N=500, numbers of communities M=4, and the new nodes is connected with two nodes in the community and one node outside the community in each time.

### 3.2 Different strategy updating rules

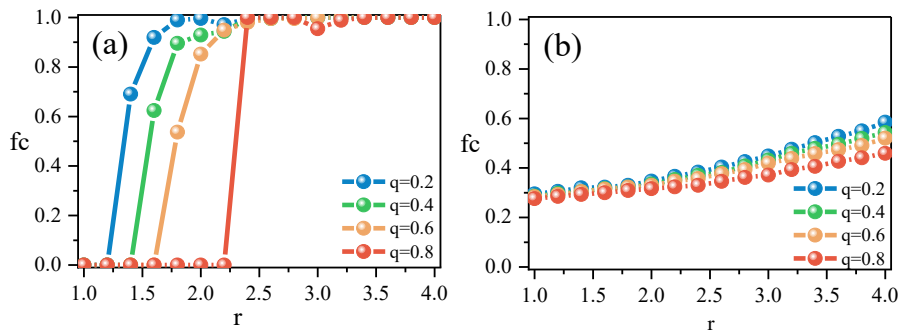

**Fig. 2.** Frequency of cooperators fc versus the enhancement factor r. (a) shows the frequency of cooperators in dependence on the parameter r when q=0.2,0.4,0.6 and 0.8 with Pairwise comparison rules; (b) shows the frequency of cooperators in dependence on the parameter r when q=0.2,0.4,0.6 and 0.8 with aspiration-driven updating rules.

Firstly, we explore the relationship between the frequency of cooperators (fc) and enhancement factor (r) under different connection probabilities when q=0.2, 0.4, 0.6 and 0.8 respectively, with respect to a single strategy updating rules in Fig.2. As shown in Fig.2, there is a positive relationship between the frequency of cooperators and the enhancement factor r. With the increase of the value of the enhancement factor, players will obtain more payoffs no matter what strategy they adopt. But obviously, the cooperators will get more benefits than the defectors. As shown in Fig.2(a), as the connection probability q increase, the phase transition from full defectors to full cooperators needs smaller time steps. But the time to phase transition gets bigger. In the real world, it is much difficult for the enhancement factor r to reach 2.5 than to reach 1.6. In other words, the probability of individuals adopting cooperation in the community network with q=0.2 is higher than that in the community network with q=0.8. Thus, it can be said that there is a negative relationship between the connection probability and the frequency of cooperators. Combining the results concluded in Fig.1. we can find that as modularity increases, the value of the frequency of cooperators gets bigger. The findings depicted in Fig.2(b) show that as r increases, the value of fc increases when q=0.2, 0.4, 0.6, and 0.8, respectively. And when the enhancement factor remains unchanged, the value of fc in the stable state of the network decreases gradually with the increase of q. It is clear that as the modularity increases, the frequency of cooperators also increases when players take aspiration-driven updating rules. We can find that the number of cooperators begins to decrease with the increase of connective probability. Thus, we can conclude that community structure can promote the emergence of cooperation.

### 3.3 Coexist updating rules

This section demonstrates the evolution of cooperation in the community network under multiple strategy updating rules. We suppose that there exist two types of players in the community network simultaneously. Individuals of type A, whose proportion is u, adopt the Pairwise comparison rules strategy updating rules, while individuals of type B, whose proportion is 1-u, adopt aspiration-driven strategy updating rules. Fig.3 and Fig.4 depict the effect of parameter u on the evolution of cooperation under different community networks. From Fig.3(a) we can find that there is a "U" shape relationship between the parameter u and fc when the connection probability is 0.2, 0.4, 0.6, and 0.8, respectively. As the parameter u increases, the frequency of cooperators decreases sightly. But once the parameter u reaches the inflection point u=0.79, with the increase of parameter u, the frequency of cooperators begins to increase rapidly. In addition, as shown in Fig.3(a), we can find that there is a negative relationship between the frequency of cooperators and the parameter q. For example, when we set the parameter u to be 0.4, the values of fc are 0.2936, 0.2691, 0.2601, and 0.2436 when q=0.2, 0.4, 0.6, and 0.8, respectively. From Fig.3(a)-(c), we can find that under different enhancement factors, the relationship among the parameter u, fc, and q remains unchanged. The only difference is that the inflection points in the relationship between the parameter u and fc changes under different values of r. As the values of r increase, inflection points get bigger.

Fig. 4 depicts the phase diagram of the frequency of cooperators in dependence on q and u. In Fig.4(a), when u is approaching to 0.7 and q is large, the frequency of cooperators is smallest. From the horizontal perspective, with the increase of u, the area of blue color first increases and then decreases with the increase of u and when u approaching to 1, the frequency of cooperators approaching to 1. From the vertical perspective, with the increase of q, the area of blue color increases. From Fig.4(a)-(c), the relationship between frequency of u and r remains unchanged, which shows that the relationship has strong robustness. Thus, we can deduce that there is a "U" relationship between the initial proportion of type A and the frequency of cooperators in public goods game and community network promotes the evolution of cooperation.

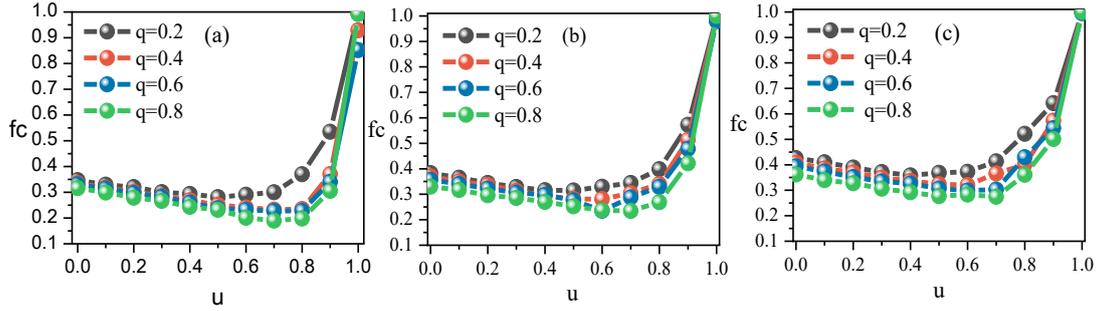

**Fig. 3.** The relationship between frequency of cooperators and initial proportion of type A u under different connection probability q. From (a) to (c), values of the enhancement factor r for different scenarios are set to be 2, 2.4, and 2.8, respectively.

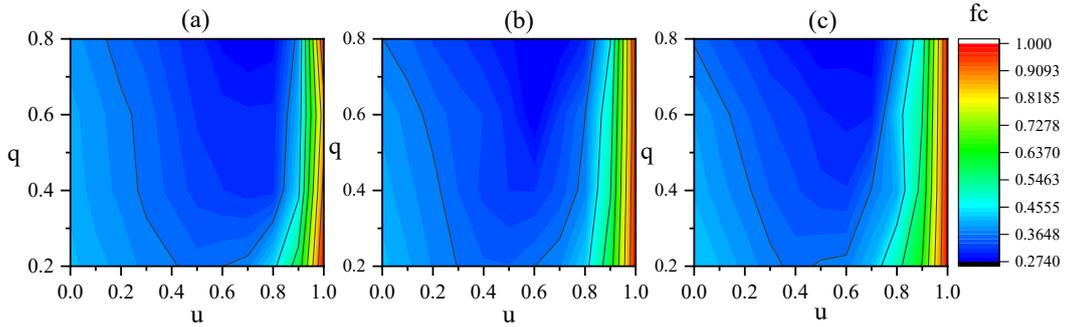

**Fig. 4.** The phase diagram of the frequency of cooperators in dependence on connection probability q and initial proportion of type A u. From (a) to (c), values of the enhancement factor r for different scenarios are set to be 2, 2.4, and 2.8, respectively.

Fig.5 and Fig.6 depict the effect of the parameter u on the evolution of cooperation under different enhancement factors. From Fig.5(a), we can find that there is a "U" shape relationship between the parameter u and the frequency of cooperators when the enhancement factor is 2.0, 2.4, and 2.8, respectively. There exists an inflection point in the relationship. When u is smaller than the inflection point, as the parameter u increases, there are no or little changes of fc. Once the value of u reaches to the inflection point, the value of fc increases rapidly with the increase of u. Moreover, the enhancement factor moderates the effect of the parameter u on the frequency of cooperators, such that the "U" shape relationship is stronger as the enhancement factor increases. When u is smaller than the inflection point, with the increase of r, the relationship between the parameter u and fc decreases faster. While when u reach to the inflection point, with the increase of r, the relationship grows faster. From Fig.5(a)-(c), we can find the similar conclusion that under different community network, the relationship among the parameter u, fc, and r remains unchanged. The only difference is that the inflection points in the relationship between the parameter u and fc changes under different values of connection probability q. As the values of q increase, inflection points get bigger.

Fig. 6 depicts the phase diagram of the frequency of cooperators in dependence on r and u. In Fig.6(a), when u is approaching to 0.72 and r is small, the frequency of cooperators is smallest. From the horizontal perspective, with the increase of u, the area of blue color first increases and then decreases with the increase of u and when u approaching to 1, the frequency of cooperators approaching to 1. From the vertical perspective, with the increase of r, the area of blue color decreases. From Fig.6(a)-(c), the relationship between frequency of cooperators, u and r remains unchanged, which shows that the relationship has strong robustness. Thus, we can deduce that there is a "U" relationship between the initial proportion of type A and the frequency of cooperators in public goods game and community network promotes the evolution of cooperation.

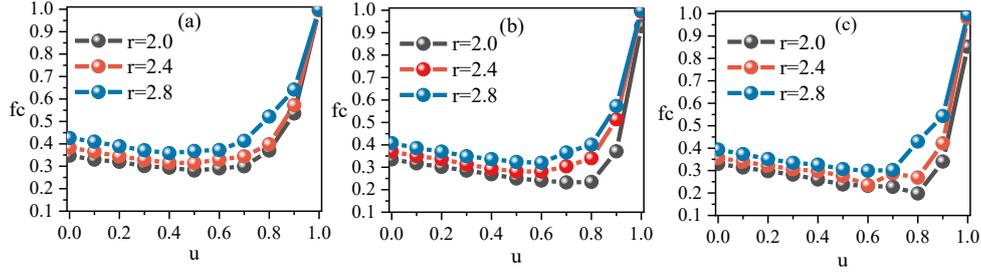

**Fig.5.** The relationship between frequency of cooperators and initial proportion of type A u under different enhancement factor r. From (a) to (c), values of the connection probability for different scenarios are set to be 0.2, 0.4, and 0.6, respectively.

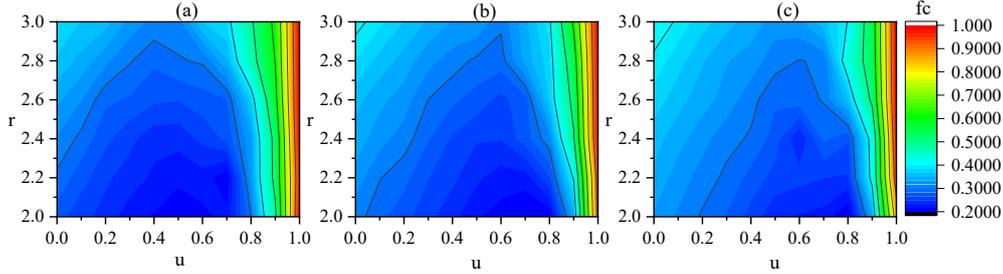

**Fig.6.** The phase diagram of the frequency of cooperators in dependence on the enhancement factor r and initial proportion of type A. From (a) to (c), values of the connection probability for different scenarios are set to be 0.2, 0.4, and 0.6, respectively.

## 4. Conclusion

This paper studies the public goods game on the scale-free network with community structure and investigates the role of heterogeneous strategy updating rules. The conclusion of this paper can be concluded as follows: (1) Community network promotes the emergence of cooperation under public goods game. When the value of modularity gets larger, the connection within communities gets closer and the connection among communities gets looser. The interact among individuals is based on public goods game, then the payoff is proportional to their investment. Once one individual took defective behavior, its payoff will more than its investment, with the evolution of time, there will be a lot of individuals even all individuals select defective behavior. Because the connection between communities is loose, individuals get their most payoff from their partners within the same communities, then the payoff of individuals will get smaller soon even there is no payoff. At this time, the individual will learn the strategy from his partner who is belong to another community, his payoff will more than his partners within the same community. Then this individual's strategy will be followed by partners within the same communities. With the evolution of time, the individuals will prefer cooperative behavior. Thus, community network benefits to the breeding of cooperative behavior.

(2) There is a positive relationship between enhancement factor and the frequency of cooperators in public goods game on the scale-free community network. Obviously, when the value of enhancement factor gets larger in the public goods game, the total payoff will increase several-fold. The payoff of individuals when they take cooperative behavior will more than the payoff when they take defective behavior, in this case individuals will be more inclined to select cooperative behavior. Thus, strengthening the enhancement factor will significantly promote the emergence of cooperation.

(3) There is a "U" shape relationship between the initial proportion of type A and the frequency of cooperators in public goods game on the scale-free community network. Taking u = 0.5 as the symmetry point, we can find that the frequency of cooperators in the group adopting Pairwise comparison rules is much higher than in the group adopting aspiration driven rules. This is because when adopting aspiration driven rules, individuals compare their payoff with the average payoff of their partners. When the individual's payoff exceeds the average payoff, it will stick to its strategy in the next round. In this case, there will always exist some individuals who adopt defection and the return is still higher than the average return. However, when adopting Pairwise comparison rules, individuals compare their own benefits with a random partner. In this case, the individual with higher payoff will be imitated by other individuals. In our experiment, the enhancement factor r>2, the group with a high proportion of cooperators creates more benefits than the group with a high proportion of defectors and then the individual who has the highest payoff is often the cooperator. Therefore, cooperators occupy a leading position in the network, and the frequency of cooperators will gradually increase.

**Acknowledgement**

The work was supported by the National Natural Science Foundation of China (No.71871182) and the Innovation Foundation for Doctor Dissertation of Northwestern Polytechnical University (No.CX2021095).

**References**

[1] J. Gore, H. Youk, A. van Oudenaarden, Snowdrift game dynamics and facultative cheating in yeast, Nature. 459 (2009) 253–256. https://doi.org/10.1038/nature07921.

[2] H.H. Lee, M.N. Molla, C.R. Cantor, J.J. Collins, Bacterial charity work leads to population-wide resistance, Nature. 467 (2010) 82-U113. https://doi.org/10.1038/nature09354.

[3] M. Milinski, TIT-FOR-TAT in sticklebacks and the evolution of cooperation, Nature. 325 (1987) 433–435. https://doi.org/10.1038/325433a0.

[4] R. Bshary, A.S. Grutter, Image scoring and cooperation in a cleaner fish mutualism, Nature. 441 (2006) 975–978. https://doi.org/10.1038/nature04755.

[5] K. Hill, Altruistic cooperation during foraging by the Ache, and the evolved human predisposition to cooperate, Hum. Nat. 13 (2002) 105–128. https://doi.org/10.1007/s12110-002-1016-3.

[6] M. Doebeli, C. Hauert, Models of cooperation based on the Prisoner's Dilemma and the Snowdrift game, Ecol. Lett. 8 (2005) 748–766. https://doi.org/10.1111/j.1461-0248.2005.00773.x.

[7] J. Zhang, F.J. Weissing, M. Cao, Fixation of competing strategies when interacting agents differ in the time scale of strategy updating, Phys. Rev. E. 94 (2016). https://doi.org/10.1103/PhysRevE.94.032407.

[8] H. Liang, M. Cao, X. Wang, Analysis and shifting of stochastically stable equilibria for evolutionary snowdrift games, Syst. Control Lett. 85 (2015) 16–22. https://doi.org/10.1016/j.sysconle.2015.08.004.

[9] M.A. Nowak, Five rules for the evolution of cooperation, Science (80-. ). 314 (2006) 1560–1563. https://doi.org/10.1126/science.1133755.

[10] M.A. Nowak, R.M. May, Evolutionary games and spatial chaos, Nature. 359 (1992) 826–829.

[11] H. Wang, Y. Sun, L. Zheng, W. Du, Y. Li, The public goods game on scale-free networks with heterogeneous investment, Phys. A Stat. Mech. Its Appl. 509 (2018) 396–404. https://doi.org/10.1016/j.physa.2018.06.033.

[12] Z. Wang, T. Chen, Y. Wang, Leadership by example promotes the emergence of cooperation in public goods game, Chaos, Solitons and Fractals. 101 (2017) 100–105. https://doi.org/10.1016/j.chaos.2017.05.027.

[13] R.R. Liu, C.X. Jia, Z. Rong, Effects of enhancement level on evolutionary public goods game with payoff aspirations, Appl. Math. Comput. 350 (2019) 242–248. https://doi.org/10.1016/j.amc.2019.01.009.

[14] M. Girvan, M.E.J. Newman, Community structure in social and biological networks, Proc. Natl. Acad. Sci. U. S. A. 99 (2002) 7821–7826. https://doi.org/10.1073/pnas.122653799.

[15] Y. Kim, S.-W. Son, H. Jeong, Finding communities in directed networks, Phys. Rev. E. 81 (2010). https://doi.org/10.1103/PhysRevE.81.016103.

[16] N. Kashtan, U. Alon, Spontaneous evolution of modularity and network motifs, Proc. Natl. Acad. Sci. U. S. A. 102 (2005) 13773–13778. https://doi.org/10.1073/pnas.0503610102.

[17] X. Bin Cao, W.B. Du, Z.H. Rong, The evolutionary public goods game on scale-free networks with heterogeneous investment, Phys. A Stat. Mech. Its Appl. 389 (2010) 1273–1280. https://doi.org/10.1016/j.physa.2009.11.044.

[18] S. Lv, J. Li, J. Mi, C. Zhao, The roles of heterogeneous investment mechanism in the public goods game on scale-free networks, Phys. Lett. Sect. A Gen. At. Solid State Phys. 384 (2020). https://doi.org/10.1016/j.physleta.2020.126343.

[19] X. Teng, S. Yan, S. Tang, S. Pei, W. Li, Z. Zheng, Individual behavior and social wealth in the spatial public goods game, Phys. A Stat. Mech. Its Appl. 402 (2014) 141–149. https://doi.org/https://doi.org/10.1016/j.physa.2014.01.064.

[20] X. Liu, Q. Pan, Y. Kang, M. He, Fixation times in evolutionary games with the Moran and Pairwise comparison rulesses, J. Theor. Biol. 387 (2015) 214–220. https://doi.org/10.1016/j.jtbi.2015.09.016.

[21] X. Liu, M. He, Y. Kang, Q. Pan, Fixation of strategies with the Moran and Pairwise comparison rulesses in evolutionary games, Phys. A Stat. Mech. Its Appl. 484 (2017) 336–344. https://doi.org/10.1016/j.physa.2017.04.154.

[22] J. Quan, Y. Zhou, X. Wang, J.B. Yang, Evidential reasoning based on imitation and aspiration information in strategy learning promotes cooperation in optional spatial public goods game, Chaos, Solitons and Fractals. 133 (2020) 109634. https://doi.org/10.1016/j.chaos.2020.109634.

[23] T. You, P. Wang, D. Jia, F. Yang, X. Cui, C. Liu, The effects of heterogeneity of updating rules on cooperation in spatial network, Appl. Math. Comput. 372 (2020). https://doi.org/10.1016/j.amc.2019.124959.

[24] T. You, L. Shi, X. Wang, M. Mengibaev, Y. Zhang, P. Zhang, The effects of aspiration under multiple strategy updating rules on cooperation in prisoner's dilemma game, Appl. Math. Comput. 394 (2021) 125770. https://doi.org/10.1016/j.amc.2020.125770.

[25] Q. Zou, K. Hu, Heterogeneous aspiration resolves social dilemma in structured populations, Chaos, Solitons and Fractals. 134 (2020). https://doi.org/10.1016/j.chaos.2020.109711.

[26] C. Li, P.K. Maini, An evolving network model with community structure, J. Phys. A. Math. Gen. 38 (2005) 9741–9749. https://doi.org/10.1088/0305-4470/38/45/002.

[27]. P. Liu, J. LIU, Cooperation in the prisoner ' s dilemma game on tunable community networks. Phys. A Stat. Mech. Its Appl. 472(2017) 156–163. http://dx.doi.org/10.1016/j.physa.2016.12.059. DOI:10.1016/j.physa.2016.12.059.

[28] B. Zhang, C. Li, Y. Tao, Evolutionary Stability and the Evolution of Cooperation on Heterogeneous Graphs, Dyn. Games Appl. 6 (2016) 567–579. https://doi.org/10.1007/s13235-015-0146-2.

[29] C. Liu, J. Shi, T. Li, J. Liu, Aspiration driven coevolution resolves social dilemmas in

networks, Appl. Math. Comput. 342 (2019) 247–254. https://doi.org/10.1016/j.amc.2018.09.034.

[30] C. Sun, C. Luo, J. Li, Aspiration-based co-evolution of cooperation with resource allocation on interdependent networks, Chaos, Solitons and Fractals. 135 (2020) 109769. https://doi.org/10.1016/j.chaos.2020.109769.

[31] J. Zhang, B. Hu, Y.J. Huang, Z.H. Deng, T. Wu, The evolution of cooperation affected by aspiration-driven updating rule in multi-games with voluntary participation, Chaos, Solitons and Fractals. 139 (2020) 110067. https://doi.org/10.1016/j.chaos.2020.110067.

[32] A. Szolnoki, J. Vukov, G. Szabó, Selection of noise level in strategy adoption for spatial social dilemmas, Phys. Rev. E. 80 (2009) 56112. https://doi.org/10.1103/PhysRevE.80.056112.

[33] X. Chen, L. Wang, Promotion of cooperation induced by appropriate payoff aspirations in a small-world networked game, Phys. Rev. E Stat. Nonlinear Soft Matter Phys. 77 (2008) 17103. https://doi.org/10.1103/PhysRevE.77.017103.

[34] M.E.J. Newman, M. Girvan, Finding and evaluating community structure in networks., Phys. Rev. E. Stat. Nonlin. Soft Matter Phys. 69 (2004) 26113. https://doi.org/10.1103/PhysRevE.69.026113.